\documentclass[aps,showpacs, pra,amssymb, amsmath, twocolumn]{revtex4}
\usepackage{amsmath}
\usepackage{amsthm}
\usepackage{eucal}
\usepackage{amssymb}
\usepackage{mathrsfs}
\usepackage{graphicx}
\usepackage{amsmath,latexsym,amssymb}
\usepackage{graphicx}
\usepackage{mathtools}
\usepackage{latexsym}
\usepackage{amssymb}
\usepackage{bm}

\begin{document}

\title{In situ single-atom array synthesis using dynamic holographic optical tweezers}

\author{Hyosub Kim, Woojun Lee, Han-gyeol Lee, Hanlae Jo, Yunheung Song and Jaewook Ahn}
\email{jwahn@kaist.ac.kr}
\affiliation{Department of Physics, KAIST, Daejeon 305-701, Korea}
\date{\today}

\maketitle

\noindent
\textbf{Laser cooling and trapping of atoms has enabled the construction and manipulation of quantum systems at the single-atom level~\cite{1,2,3,4,5,6,7,8}. To create scalable and highly controllable quantum systems, e.g., a large-scale quantum information machine, further development of this bottom-up approach is necessary. The implementation of these systems requires crucial prerequisites: scalability, site distinguishability, and reliable single-atom loading onto sites. The previously considered methods~\cite{9,10,11,12} satisfy the two former conditions relatively well; however,  the last condition, loading single atoms onto individual sites, relies mostly on probabilistic loading~\cite{2,6,9,10,11,12}, implying that loading a predefined set of atoms at given positions will be hampered exponentially. Two approaches are readily thinkable to overcome this issue: increasing the single-atom loading efficiency~\cite{13,13-1,14} and relocating abundant atoms to unfilled positions~\cite{15}. Realizing the atom relocation idea, in particular, is directly related to how many atoms can be transportable independently. Here, we demonstrate a dynamic holographic single-atom tweezer with unprecedented degrees of freedom of $2N$. In a proof-of-principle experiment conducted with cold $^{87}$Rb atoms, simultaneous rearrangements of $N=9$ single atoms were successfully performed. This method may be further applicable to deterministic $N$ single-atom loading, coherent transport~\cite{16,17}, and controlled collisions~\cite{18,18-1}. }
 
The advantage of using holographic optical tweezers is that arbitrary potentials can be designed for atoms~\cite{6,11}. Because the optical tweezers are the image determined by the wave propagation integral, e.g., Fourier transformation (FT), the hologram for a complex potential can be designed using a numerical method often based on iterative FT algorithms (IFTAs)~\cite{11}. When being used in conjunction with an active holographic device such as spatial light modulators (SLM), IFTA can produce dynamic optical potentials, of which the many applications include dynamic in situ atom manipulation~\cite{16,17,18}, quantum logic gate~\cite{19}, pattern formations in an addressable optical lattice~\cite{15}, and real-time feedback transportation of atoms~\cite{7}. This combination, however, has yet to show reliable performance with the exception of a few trivial demonstrations on a small scale~\cite{20}. 

\begin{figure}[htb]
\centering
\includegraphics[width=0.48\textwidth]{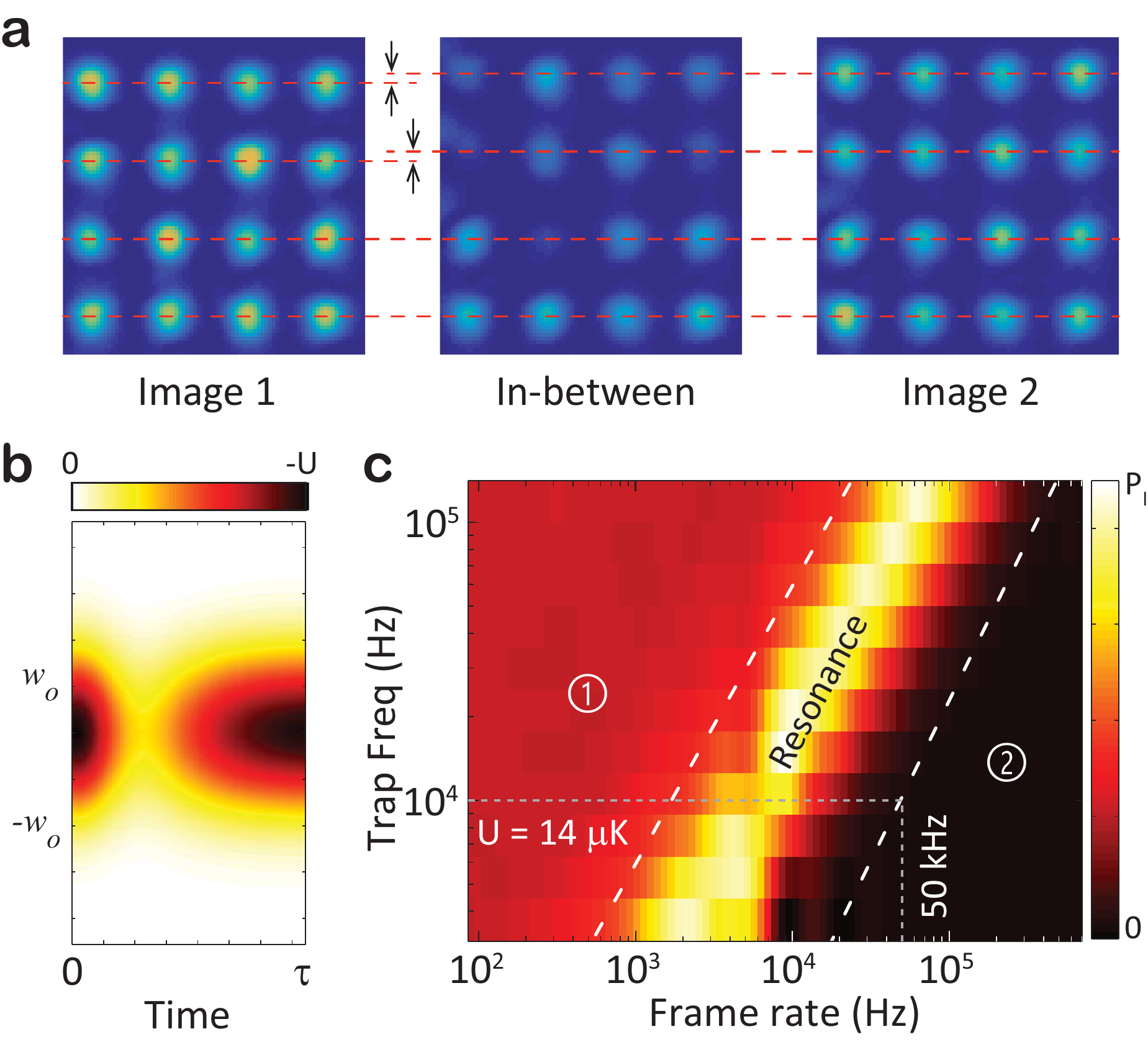} 
\caption{\textbf{Intensity flicker of IFTA transport.} \textbf{a}, Stroboscopic measurement of optical images of the trap array. Two different images are generated using an IFTA and liquid crystal SLM (LCSLM).
\textbf{b}, The transient potential for the trap loss simulation. The trap waist is $w_o=1.14$~$\mu$m, the transient time $\tau=1/f$ where $f$ is frame rate, and the displacement $w_0/18$. The color scale is normalized by the peak potential, $–U$. 
\textbf{c}, Trap loss landscape by the flickering potential (\textbf{b}).  The color scale normalized by $P_{l}$ represents the loss probability at time $\tau$ which varies by $0.005-0.04$ according to the initial trap condition ($T/U=1/18 \sim 1/12$).
}
\label{fig1}
\end{figure}

Despite their promising utility, active holographic optical tweezers are unable to sustain trapped atoms while the hologram is being updated because of intensity flickering~\cite{21}. Even if individual holograms generated by an IFTA form the required optical potentials, the frame-to-frame evolution does not necessarily maintain an appropriate in-between potential because cross-talk noises occur ({see Fig.~\ref{fig1}a}). Often such intensity flickering is significant and irregular over the entire range of the optical potential, and intensity feedback control is not effective for such an irregular transient. 
While the intensity flickering is not an issue for macroscopic particles suspended in a solution~\cite{22}, microscopic particles (e.g., atoms) do not wait until the missing potential recovers or cannot resist excessive displacement heating. Even with a fast device such as a digital micromirror device (50-kHz frame)~\cite{23} or for ultracold atoms~\cite{24}, a large portion of the trapped atoms is lost. 
The trap-loss simulation ({see Methods}) performed as a function of the frame rate, $f$, of the device and the trap frequency, $f_r=1/2\pi\sqrt{4U/mw^2_o}$, supports  the idea that the intensity flickering hinders the trap stability ({see Figs.~\ref{fig1}b and \ref{fig1}c}). 
In particular, a constant loss exists because of the significant intensity flickering in the adiabatic region ($f_r \gg f$, Region~\raisebox{.5pt}{\textcircled{\raisebox{-.9pt} {1}}} in {Fig.~\ref{fig1}c}). In the non-adiabatic region ($f_r<f$, Region~\raisebox{.5pt}{\textcircled{\raisebox{-.9pt} {2}}} in {Fig.~\ref{fig1}c}), single steps do not lose atoms; however, in this region, either the atoms boil up fast via displacement heating or current technologies are not applicable. 
The macroscopic particle is a schematic asymptote of $f_r\rightarrow 0$ as $m \rightarrow \infty$, and the suspending solution is a heat bath that rapidly dissipates the excessive energy from the displacement heating; thus, movable conditions can be directly achieved even with the significant intensity flicker~\cite{22}. The holographic transport of single atoms, however, requires an alternative algorithmic approach. 

\begin{figure*}[htb]
\centering
\includegraphics[width=0.95\textwidth]{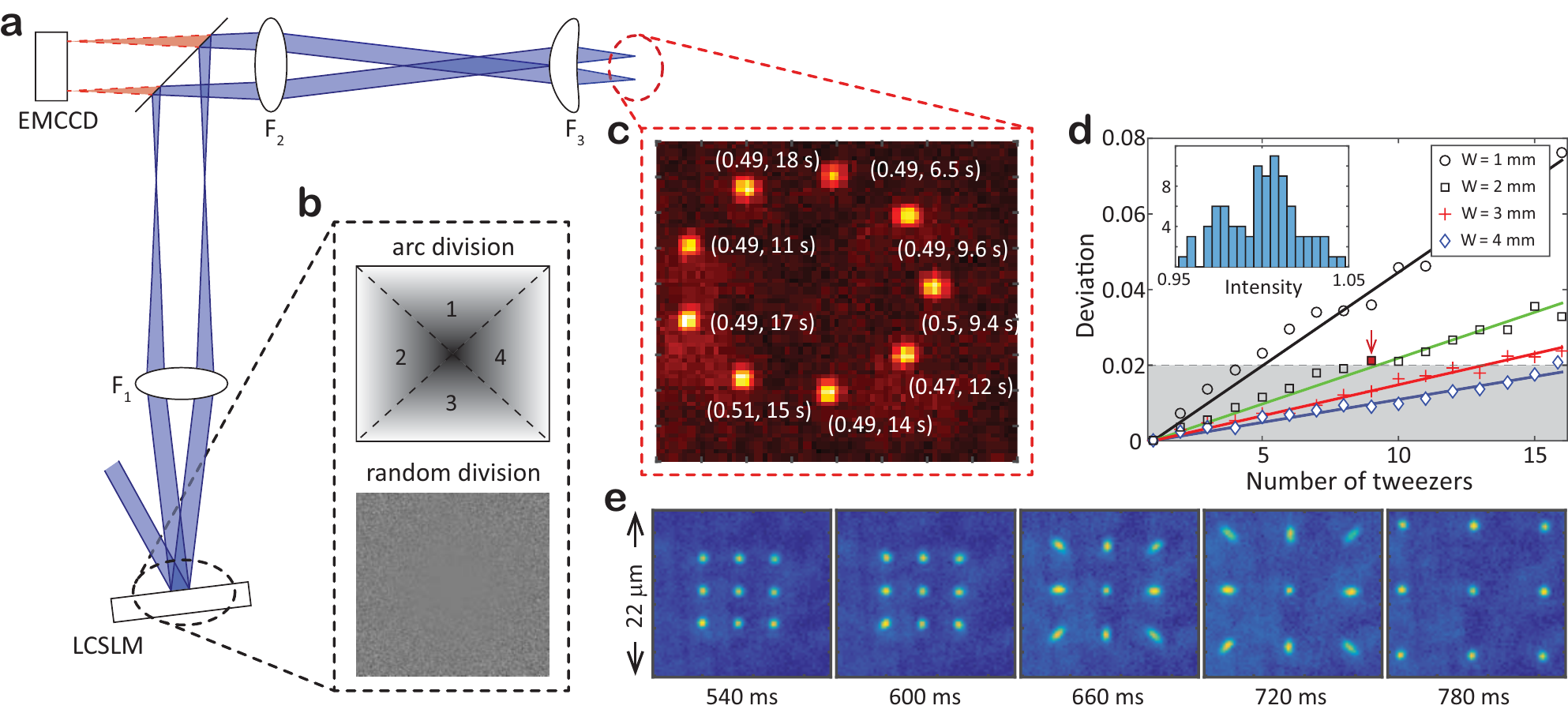}
\caption{\textbf{Set-up for single-atom holographic transport.} \textbf{a}, The optical system for hologram transfer and trap imaging. \textbf{b}, Schematic 2D phase planes of the LCSLM in gray scale. One of them intuitively illustrates the working principle (arc division), and the other shows real hologram (random division), respectively. 
 \textbf{c},  An example of a $N=9$ loaded single-atom array, represented with $22 \times 22$~$\mu$m$^2$ size 500 cumulative images. The parenthesis denotes the loading probability and life time. \textbf{d}, The intensity standard deviation as a function of $N$ for the random division. The data points were calculated for various beam waists $W$ at the SLM window, and the solid lines are the linear fits to the data. The slopes are precisely proportional to $1/W$. The inset presents the intensity histogram of $W=2$~mm for $N=9$ of the red mark. \textbf{e}, An in situ single-atom array expansion movie.}
\label{fig2}
\end{figure*}

Our algorithm uses the simplest analytic form of beam steering, {\it i.e.}, the linear phase $\phi(x)=k_xx$~\cite{24-1}. This phase modulation directly couples the modulation plane (Fourier domain) control parameter, $k_x$, to the image plane position, $X=Fk_x/k$, with one-to-one correspondence, where $F$ is the lens focal length and $k$ is the wave-vector of the trap light. The given function is a flicker-free solution, because a linear combination of two phases, $\alpha k_1x+(1-\alpha)k_2x$ with $\alpha \in [0, 1]$, is again a linear phase which smoothly sweeps the two focal points, $X_1=Fk_1/k$ and $X_2=Fk_2/k$. To obtain more than a single optical tweezer, the modulation plane is divided into several sub-planes (see Fig.~\ref{fig2}b). When each sub-plane is assigned to each linear phase and the division is randomized in the single-pixel resolution of the device, this method effectively preserves the diffraction limit of the individual optical tweezers focused onto the image plane. In this manner, the required trap laser power scales with $N^2$, where $N$ is the number of optical tweezers. It is inefficient compared to the IFTA, which scales with $N$, because the IFTA coherently sums over the entire modulation plane for every optical tweezer. Nevertheless, the proposed algorithm concedes power efficiency for independent controllability without intensity flickering.

Figure~\ref{fig2}a depicts the set-up consisting of an active holographic device (LCSLM), an imaging system, and a cold $^{87}$Rb atom chamber (not shown). The holograms were transferred by a two-lens ($F_1$, $F_2$) system to the entrance pupil of a high numerical aperture ($\rm {NA}=0.5$) lens ($F_3$)~\cite{11}. The optical tweezers ($\lambda=820$~nm) on the final image plane had a beam radius of $w_o= 1.14$~$\mu$m, a trap depth of $U=1.4$~mK, and an optical power of $P_o=3.4$~mW per optical tweezer. The trap frequencies were $f_r=100$~kHz and $f_z=17$~kHz for the radial and axial directions, respectively. The 3D molasses continuous imaging~\cite{9,25} captured a snapshot every 60~ms. An image accumulating 500 snapshots is shown in Fig.~\ref{fig2}c. The temperature of the trapped atoms was measured as $T=80 \sim 140$~$\mu$K using the release and recapture method~\cite{26}, where the error was the 1$\sigma$ band of each optical tweezer. 
As $N$ increases, the number of SLM pixels per single optical tweezer decreases, thus degrading the optical tweezer shape and intensity regularity (Fig.~\ref{fig2}d). To maintain an acceptable quality of the optical tweezers for the experiments, we chose the maximum number  $N=9$ for $W=2$~mm. The quality factor, the standard deviation of the individual peak intensity, was less than 0.02. 
Thus, the LCSLM in our current setup would support up to $N=17$ optical tweezers in the given standard. For the required power scaling as  $N^2$, the number of available optical tweezers scales as $W$. Therefore, the SLM damage is not an issue because the required intensity onto the SLM is preserved.
Figure~\ref{fig2}e shows an array-rearrangement demonstration based on our algorithm. The series of images each accumulating 500 snapshots of the experiment demonstrates that the initially prepared 3-by-3 square array of single atoms with a spacing of $d=4.5$~$\mu$m expands to an array of twice the lattice spacing of $2d$. Because each atom moves in 2D with parameters $(X_i, Y_i)$ for $i=1$ to $N=9$, the total degrees of freedom of movement are $2N=18$. 

\begin{figure}[htb]
\centering
\includegraphics[width=0.48\textwidth]{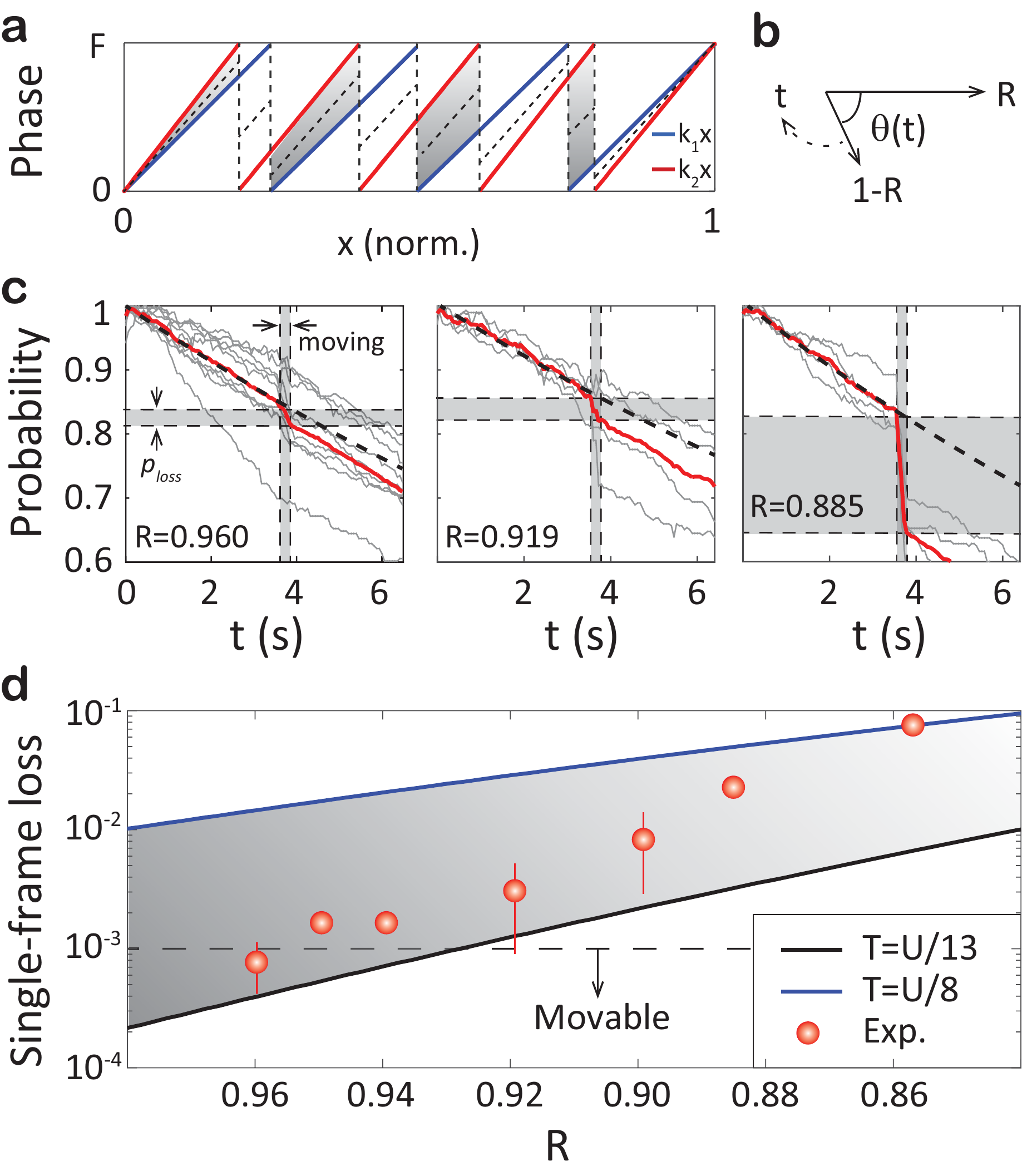} 
\caption{\textbf{Transport loss mechanism.} \textbf{a}, Schematic modulated two linear phases. The $x$-axis is normalized to 1. The sum of the shaded regions are the flicker-free ratio $R$ which is proportional to field strength. \textbf{b}, Transient vectorial representation of the flicker effect. \textbf{c}, The gray dotted lines represent the individual trap decay curves, and the red solid line is averaged over them. The black dashed line is an exponential fit to the red solid line.  The $x$-axis is the time-lapse after the array loading, and the  $y$-axis is the single-atom survival probability measured using 1,000 cumulative events and normalized to 1. \textbf{d}, Single-frame losses  measured for various $R$s. The red circles are extracted single-frame data from \textbf{c} and the error bars represent the standard deviations of the different sites. The solid lines are from the adiabatic intensity flickering model.}
\label{fig3}
\end{figure}

The feature of the LCSLM most strongly coupled to the intensity flicker is the finite modulation depth $\Phi$ ($=2\pi$). 
When a linear phase gradually changes from $k_1x$ to $k_2x$, as depicted in Fig.~\ref{fig3}a, certain regions (shaded) are flicker-free, but the rest are not due to the $\Phi$ phase jump. In Fig.~\ref{fig3}b, $R$ (the flicker-free ratio, see Methods) denotes the vector sum of fields from the all shaded regions and $1-R$ the field from the rest regions, when the field sum from the all regions is normalized. Then, the peak intensity of an optical tweezer varies between 1 and $(2R-1)^2$ during the phase evolution from $k_1x$ to $k_2x$, and the intensity flicker can be quantified by $4R(1-R)$. In Fig.~\ref{fig3}c, the atom decay curves are measured under the influence of the intensity flicker (the vertical shaded region denoted by moving), where the intensity flicker was induced by the 5-$\mu$m displacement of the optical tweezer. The sudden probability drop, $p_{\rm loss}$, is measured for each $R$ value, and, from the measurement, the single-frame loss probability, $P_{\rm l}=1-(1-p_{\rm loss})^{1/n}$, where $n$ is the number of frames in each displacement, is obtained as in Fig.~\ref{fig3}d. When the result is compared with the adiabatic intensity flickering model (see Methods), the measured atom temperature agrees well with the temperatures of the two theory lines, 107 and 175~$\mu$K, respectively, in Fig.~\ref{fig3}d. The agreement supports that the dominant moving loss mechanism is the intensity flicker. Because the theory predicts that the single-frame loss can be exponentially decreased as a function of $R$, the high fidelity holographic transport is achievable. For example, when we consider an empirically defined movable region of $R>0.96$, which is about 200-nm step, the fidelity after a 5-$\mu$m transport exceeds $\sim$0.99. 

\begin{figure}[htb]
\centering
\includegraphics[width=0.48\textwidth]{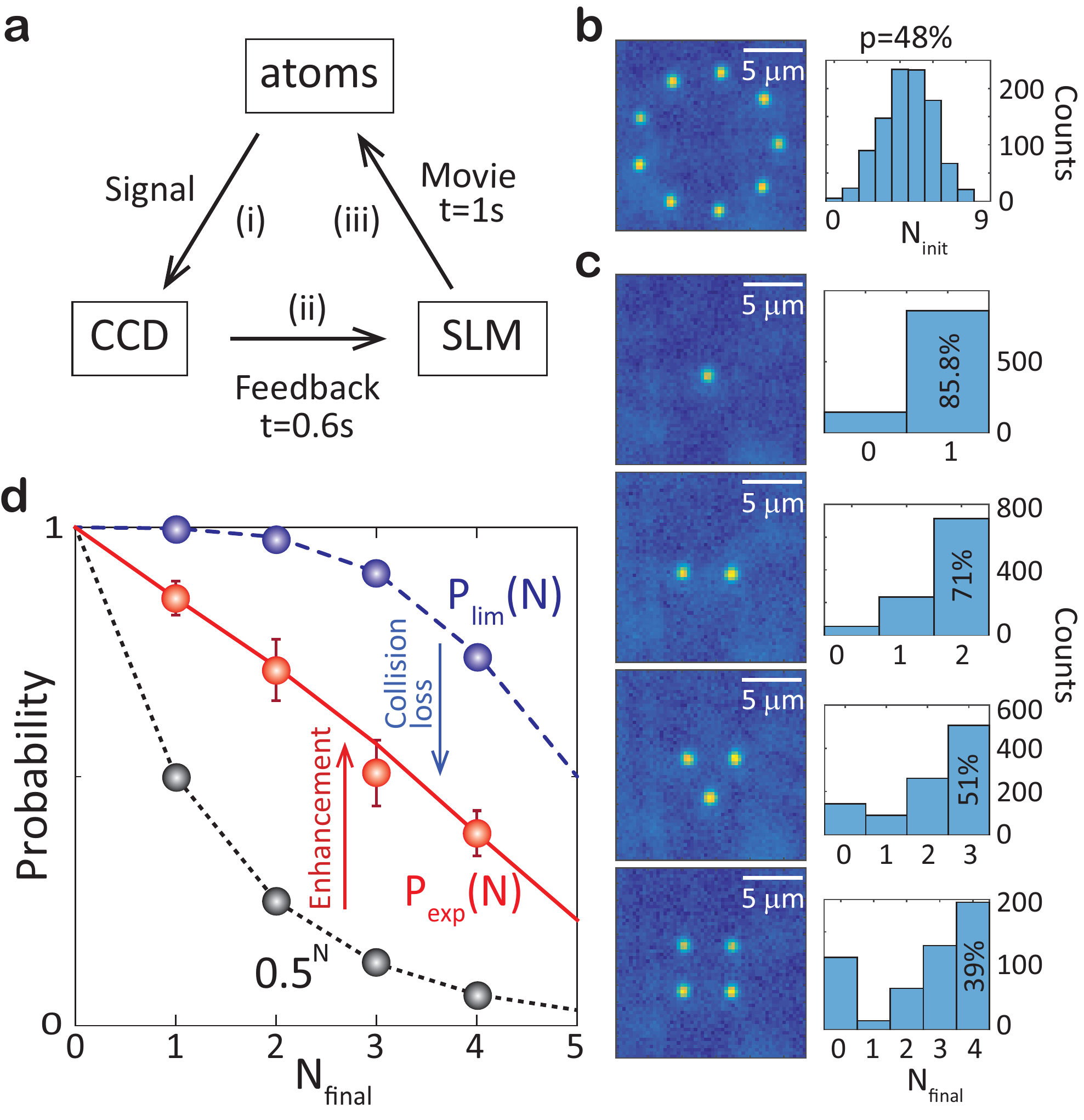} 
\caption{\textbf{In situ single-atom array synthesis.} \textbf{a}, The feedback-control loop sequence: (i) signal gathering and processing, (ii)  state resolving and solution finding, (iii) solution execution. \textbf{b}, The cumulative image of initially loaded $N=9$ atoms (left), accompanied by the corresponding number histogram (right). \textbf{c}, From the initial 9 sites, one, two, three, and four atoms are rearranged through single feedback loop. \textbf{d}, Success probabilities of the atom rearrangement: $p=0.5$ (black dotted line), experiments (red line and data points), and theoretical limit (blue dashed line). The error bar depicts the standard deviation of time-binned 100 events. }
\label{fig4}
\end{figure}

Finally, Fig.~\ref{fig4} presents a proof-of-principle demonstration of the in situ single-atom array synthesis. In a three-step feedback loop of initial atom loading in $N_{\rm init}=9$ sites, read-out, and rearrangement as shown in Fig.~\ref{fig4}a, an atom array of $N_{\rm final}=1$, 2, 3, or 4, is produced. After the atoms are initially loaded at the sites, the first computer checks the occupancy of each site by reading out the electron multiplying charge-coupled device (EMCCD) images. A 9-bit binary information that represents the occupancy is then sent to the second computer, which has a look-up library of atom rearrangement trajectories, each stored in DRAM as a sequence of 30 holograms,
between all possible initial and final pairs of atom arrays. 
It takes 0.6 seconds from the read-out to the retrieval of an appropriate trajectory. Then, the second computer sequentially loads the holograms to the SLM at a speed of 30 fps to move the atoms along the trajectory. The initial, and four types of final cumulative images are represented by the atom number histograms in Figs.~\ref{fig4}b and \ref{fig4}c. The individual final images are independent experiments that form one, two, three, or four atom arrays out of nine. 

Compared with the binomial distribution of the initial histogram, the final histograms have highly unconventional distributions. The loading efficiency curves of the final arrays are presented in Fig.~\ref{fig4}d, where the red  points are the data from the number histograms of at least 500 events. The black dotted line follows $0.5^N$, which is the loading efficiency in the collisional blockade regime. The blue dashed line follows the cumulative binomial distribution given by
\begin{equation}
P_{\rm lim}(N)=\sum_{n=N}^{9} \binom {9} {n}p^n(1-p)^{9-n},
\end{equation} 
and the red line is 
\begin{equation}
P_{\rm exp}(N)=P_{\rm lim}(N)\times p_s^N,
\end{equation}
where $p=0.48$ is the initial loading probability and $p_s=0.86$ is the experimental (moving) success probability from the fitting. $1-p_s$ is composed of the background collisional and moving losses. During the entire feedback process, the background collisional loss is estimated to be 0.13 and the moving loss is estimated to be $0.01$. The moving process, as expected, has high fidelity that a deterministic transport has been reliably completed. The $N_{final}=4$ case exhibits six-fold enhanced loading efficiency compared with the $p=0.5$ collisional blockade regime.

Our method can be further improved by increasing the number of atoms and the efficiency of initial loading. Besides simply increasing the number of dynamic holographic tweezers, we may also use a passive diffractive optical element (DOE). With the commonly available 6-by-6 passive DOE,  for example, the probability for at least 9 atoms to be initially captured out of 36 sites can exceed $99.8\%$ at $p=0.48$. In addition, when the background collision is minimized in a low-pressure chamber of low $10^{-11}$~Torr, $p_s$ can be minimized to as low as $p_s= 0.97$ to increase the probability of creating a completely packed 3-by-3 atom array to as high as $80\%$. 
Control system optimization would assist in further improvement, e.g., from two-computer TCP:IP communication to a single-computer integrated system and a 60-fps movie would at least double the feedback loop speed. The speed of the liquid-crystal dynamics ($<$100 Hz) limits the performance of the given method but the fundamental limit of the dynamics would be much faster~\cite{29}. 
   
In summary, the analytic approach to optical potential design has demonstrated the holographic transport of single-atom arrays. The systematic analysis of intensity flicker enabled  the moving loss to be parameterized; thus, we could find and achieve the deterministic transport regime using a holographic method. An individual atom has its own degrees of freedom in the image plane; thus, a total moving degrees of freedom of $2N$ was achieved, which represents unprecedented space controllability. Furthermore, we could see that the overlapping Fresnel lens pattern (not shown here) can transport the array in the axial direction, suggesting that $3N$ degrees of freedom is also possible, but remains to be investigated. We also formed an in situ feedback loop for atom array rearrangement, which is a proof-of-principle demonstration of high-fidelity atom array preparation. Its possible application is not limited to the array preparation but may be applied to many-body physics with arranged atoms~\cite{18-1} and coherent qubit transports~\cite{17,18}. 
 
 \begin{acknowledgements}
This research was supported by the Samsung Science and Technology Foundation [SSTF-BA1301-12]. The construction of the cold atom apparatus was in part supported by the Basic Science Research Program [2013R1A2A2A05005187] through the National Research Foundation of Korea.

\textbf{Authors Contributions}  H.~K., H.~L., Y.~S., and H.~J. constructed the experimental apparatus; H.~K. and W.~L. took the data and performed the data analysis, with guidance from J.~A. provided theoretical support. All authors contributed to the writing of the manuscript.

\end{acknowledgements}

\pagebreak
\begin{center}
\textbf{\large METHODS}
\end{center}
\setcounter{equation}{0}
\setcounter{figure}{0}
\setcounter{table}{0}
\setcounter{page}{1}
\makeatletter
\renewcommand{\theequation}{S\arabic{equation}}
\renewcommand{\thefigure}{S\arabic{figure}}
\renewcommand{\bibnumfmt}[1]{[S#1]}
\renewcommand{\citenumfont}[1]{S#1}


\noindent \textbf{Experiments} The active holographic device was a liquid crystal SLM (HOLOEYE, PLUTO), a reflective phase modulator array of $1920 \times 1080$ pixels with an 8~$\mu$m$^2$ pixel size and the first-order diffraction efficiency was $\sim$50\%.
The far-off-resonant trap (FORT) beam of $P_{\rm total}=1.1$~W from a Ti:sapphire continuous-wave laser (M SQARED, SolsTiS) was tuned at $820$~nm  to illuminate the SLM with a beam radius of $W=2~$mm in a near-orthogonal incident angle.
The diffracted beam from the SLM was imaged onto the intermediate image plane by an $F_1=200$~mm lens, and then re-imaged onto the entrance pupil of the objective lens by a second lens (doublet, $F_2=200$~mm).
The objective lens (Mitutoyo, G Plan Apo) was an infinity-corrected system, having a focal length of $F_3=4$~mm, a numerical aperture of $\rm{NA}=0.5$, and a long working distance of $16$~mm with 3.5~mm-thick glass-plate compensation. 
Then the given laser power was able to sustain up to $N=\sqrt{P_{\rm total}\eta_d \eta_s /P_o}=9$ optical tweezers, where the diffraction efficiency was $\eta_d=0.5$, the optical system loss $\eta_s=0.5$, and $P_o=3.4$~mW.
 
The cold $^{87}$Rb atom chamber was a dilute vapor glass cell in a constant pressure of $3\times 10^{-10}$~Torr. It had four $100 \times 40$~mm$^2$ clear windows with a thickness of $3.5$~mm. The $^{87}$Rb atoms from a getter were captured by a six-arm 3D magneto-optical trap (MOT) with a beam diameter of 7.5~mm~$(1/e^2)$, a detuning of $-18$~MHz from the $5S_{1/2} (F=2) \rightarrow~5P_{3/2} (F'=3)$ hyperfine transition, and $dB/dz=15G/cm$.
 After an initial MOT loading operation for 2.8~s, the atom density became $\sim 10^{10}~$cm$^{-3}$ (equivalent to 0.2 atoms per single trap volume), so the $-46$-MHz detuned 3D molasses and the FORT were overlapped for 200~ms to achieve the collisional blockade regime of the $p=0.5$ filling probability in every site~[2]. 
After this, the magnetic-field gradient and the molasses were turned off for $100$~ms to dissipate residual cold atoms and then the 3-D molasses were turned back on for continuous imaging~[9,28].
The scattered photons were collected by the same objective lens $F_3$ and imaged onto the EMCCD (Andor, iXon3 897) through the lens $F_2$ with an overall efficiency of $\eta_c=0.02$.
The image plane of 26$\times$26~$\mu$m$^2$ was captured as a snapshot in every 60~ms.
The trap lifetime was measured as 12~seconds, consistent with the effect of the background gas collision.

\noindent \textbf{Single-atom detection} 
The scattering cross-section is given by 
$\sigma={\sigma_0}/[{1+4(\Delta/\Gamma)^2+I/I_{\rm sat}}]$,
where the resonant cross-section is $\sigma_0=2.907\times 10^{-9}$~cm$^2$, the natural line width $\Gamma=5.746$~MHz, the saturation intensity $I_{\rm sat}=1.669$~mW/cm$^2$, the detuning $\Delta=-100$~MHz (Stark shift is considered), and the intensity $I=27$~mW/cm$^2$. Each atom in an optical tweezer emits $2.87\times 10^5$/s photons. With the overall detection efficiency of 0.02 and the exposure time of 50~ms, we collect 280 photons per atom. It corresponds to a signal-to-noise ratio (SNR) $>5$ for the EMCCD. When a Gaussian noise is assumed, the theoretical discrimination probability between zero and single atom is given with $5\sigma$ significance or 99.99995\%. In the experiments, a background photon noise exists, but the histogram shows the success probability exceeding 99.99\%.

\noindent \textbf{Heating in optical tweezers} 
There are several heating sources for trapped atoms including the FORT scattering, the intensity noise, the beam pointing fluctuation;  however, none of them are a significant heating source. The heating caused by the photon scattering from FORT is estimated as 7~$\mu$K/s at the peak intensity, which is negligible in our trap. The intensity fluctuation ($\Delta I$) and beam pointing fluctuation ($\Delta \omega_o$) by the LCSLM voltage update are up to 6\% and 5\%, respectively; its noise spectrum, however, is less than 500~Hz. The parametric heating has harmonic resonances at $\omega_r/2n$~\cite{s1, s2}, which are far from 500~Hz. The quantitative heating rate has not been estimated; however, any atom loss difference is not observed during the one second transport without the molasses. Empirically the low frequency noise does not degrade the trap capability in our 1.4~mK deep trap. Utilizing an intensity feedback control to the diffraction beam will diminish the intensity fluctuation~\cite{s0}; thus, a lower trap depth would be achievable.

\noindent \textbf{Dynamic range and resolution of the control space}
The optical tweezers are separated from the zeroth order diffraction by $X_{01}=F_1F_3/F_2\times k_1/k>5$~$\mu$m to avoid the cross-talk, which sets the lower limit $X_{01}^{\rm min}=5$~$\mu$m of the dynamic range of the control space. The upper limit is empirically given by $X_{01}^{\rm max}=45$~$\mu$m, because the diffraction efficiency decreases as $k_1$ increases. Note that the entrance pupil diameter of the objective lens is $D\sim2 {\rm NA} F_3$ and the initial beam diameter $2W=4$~mm at the SLM is (de)magnified by the ratio $F_2/F_1=1$ to fit with $D$ for optimal performance, which results in $X_{01}=D/2{\rm NA}\times k_1/k$, independent of the focal lengths of the system. In our experiment, a safe working area of 26$\times$26~$\mu$m$^2$ is used for the optical tweezer patterns and the imaging plane. The discrete phase induces a beam steering error~[27] but the amount of the error is only 4~nm in our system with 256 gray levels. Thus the resolution is limited by 4~nm, which is much smaller than the long term drift (100~nm) and the LCSLM refresh fluctuation (100~nm).

\noindent \textbf{Adiabatic intensity flicker model}
The probability for an atom initially trapped in a potential $U$ to escape from an adiabatically lowered potential $U'$ is approximately given by
\begin{equation}
P_{l}=\int^{\infty}_{U'}\frac{E^2}{2 (k_b T')^3}e^{-\frac{E}{k_b T'}}dE,
\label{eqPl}
\end{equation}
where $T'=T\sqrt{U'/U}$ is the temperature of the Boltzman distribution~[29]. In our experiment, the lowest trap potential is given by $U'=(2R-1)^2U$, where $R$ is the flicker-free ratio. The solid lines in Fig.~\ref{fig3}d are the numerically obtained results of Eq.~\ref{eqPl} for $T=U/13$ and $T=U/8$, respectively.

\noindent \textbf{Flicker-free ratio $R$}
The finite modulation depth ($ 0 \leq \phi \leq \Phi$) of the SLM phase restricts the ideal linear phase to a modulated phase in a sawtooth shape; thus, $R$ is directly related to $\Phi$. We consider two SLM phases $\phi_1(x)=mod(k_1x+\Phi/2,\Phi)$ and $\phi_2(x)=mod(k_2x+\Phi/2,\Phi)$, where $\Phi \in 2\pi N$, $x \in [-D/2, D/2]$, $D=4.6$~mm is the size of the active SLM window, and we assume $k_1 \leq k_2$ without loss of generality. The phase evolution from $\phi_1(x)$ to $\phi_2(x)$ is then given by $\phi(x,t)=\phi_1(x)e^{-t/\tau}+\phi_2(x)(1-e^{-t/\tau})=k_1xe^{-t/\tau}+k_2x(1-e^{-t/\tau})-\Phi N_1(x)e^{-t/\tau}-\Phi N_2(x)(1-e^{-t/\tau})$, where $N_{1,2}(x)=[ k_{1,2}x/\Phi ]$ is a function defined with the Gauss' symbol $[x]=x-mod(x)$ and $\tau$ denotes the response time. The condition $N_1(x)=N_2(x)$ defines the flicker-free evolution regions (the shaded regions in Fig.~\ref{fig3}a). In our experiment, the flicker-free regions are divided into two regions respectively satisfying $N_1(x)=N_2(x)$ and $N_1(x)=N_2(x)-1$, because $R$ is large enough for an atom transport or $(k_2-k_1)(D/2) < 2\Phi$.  As a result, the flicker-free ratio $R$ defined by the sum of the flicker-free regions divided by $D/2$ is given by
\begin{eqnarray}
R &= & \frac{2\Phi}{D} \sum_{n=1}^{N_2^{f}} \Big\{ \frac{n}{k_2}-\frac{(n-1)}{k_1} \Big\}+  1-\frac{2\Phi}{k_1 D}N_2^{f} 
\nonumber \\
&=& \frac{\Phi}{k_2D}N_2^{f} (N_2^{f} +1)-\frac{\Phi}{k_1D}N_2^{ f} (N_2^{ f} +1)+1
\label{eqR}
\end{eqnarray}
for the case of $N_1({D}/{2})=N_2({D}/{2})\equiv N_2^f$ and 
\begin{eqnarray}
R &= & \frac{2\Phi}{D} \sum_{n=1}^{N_2^{f}} \Big\{ \frac{n}{k_2}-\frac{(n-1)}{k_1} \Big\} \nonumber \\
&=& \frac{\Phi}{k_2D}N_2^f(N_2^f+1)-\frac{\Phi}{k_1D}N_2^f(N_2^f-1)
\end{eqnarray}
for the case of $N_1({D}/{2})\neq N_2({D}/{2})$.
The $R$ in Eq.~\eqref{eqR} can be further simplified to
\begin{equation}
R=1-\frac{{k_2D}+2\Phi}{4\Phi}\Big(1-\frac{k_1}{k_2}\Big)
\label{eqR2}
\end{equation}
under the assumption $N_1=N_2\approx{k_2D/2/\Phi}$. The obtained analytic result for the $R$ is within 2\% difference from the actual numerical value estimated by considering the circular active SLM window, the Gaussian beam profile of the optical tweezers, the 2D nature of the $k1$ and $k2$, and the discrete pixel size of the SLM. Within the experimental variation of $k1$ and $k2$, $R$ varies from 0.86 to 0.96 as shown in Fig.~\ref{fig3}d.

\noindent \textbf{Single-frame displacement vs. $R$}
The maximum single-frame displacement $\Delta X_{01}=\frac{D}{2k NA}\Delta k$ is also given as a function of $R$. 
When Eq.~\eqref{eqR2} is expressed with $k_2D=2k{\rm NA} X_{01}$ and $1-k_1/k_2=\Delta X_{01}/X_{01}$, we obtain
\begin{equation}
\frac{\Delta X_{01}}{X_{01}} \leq \frac{2(1-R)\Phi}{2{\rm NA}kX_{01}+\Phi},
\label{restriction}
\end{equation}
so the single-frame displacement is proportional to $1-R$.

\noindent \textbf{The trap loss simulation in Figs.~\ref{fig1}b and \ref{fig1}c} For a pair of initial and final trap potentials, which have $N$ trap sites, two phase holograms are calculated using Gerschberg-Saxton algorithm, respectively. Some of the sites in the initial trap potential are separated by $1/18w_o$ from the final trap potential. The in-between holograms are constructed by $\phi_1e^{-t/\tau}+\phi_2(1-e^{-t/\tau})$, which generate the transient behavior of a single trap potential as shown in Fig.~\ref{fig1}b. Then, the trajectories $(p, q)$ of the 1D classical Hamiltonian equation of motion calculated by the symplectic Euler method are used to estimate the trap loss probability in Fig.~\ref{fig1}c, where we use a loss criteria of $|q(t)| > 2.5w_o$ and the initial energy and positions are sampled using the Monte-Carlo method~[29].

\begin{figure}[htb]
\centering
\includegraphics[width=0.45\textwidth]{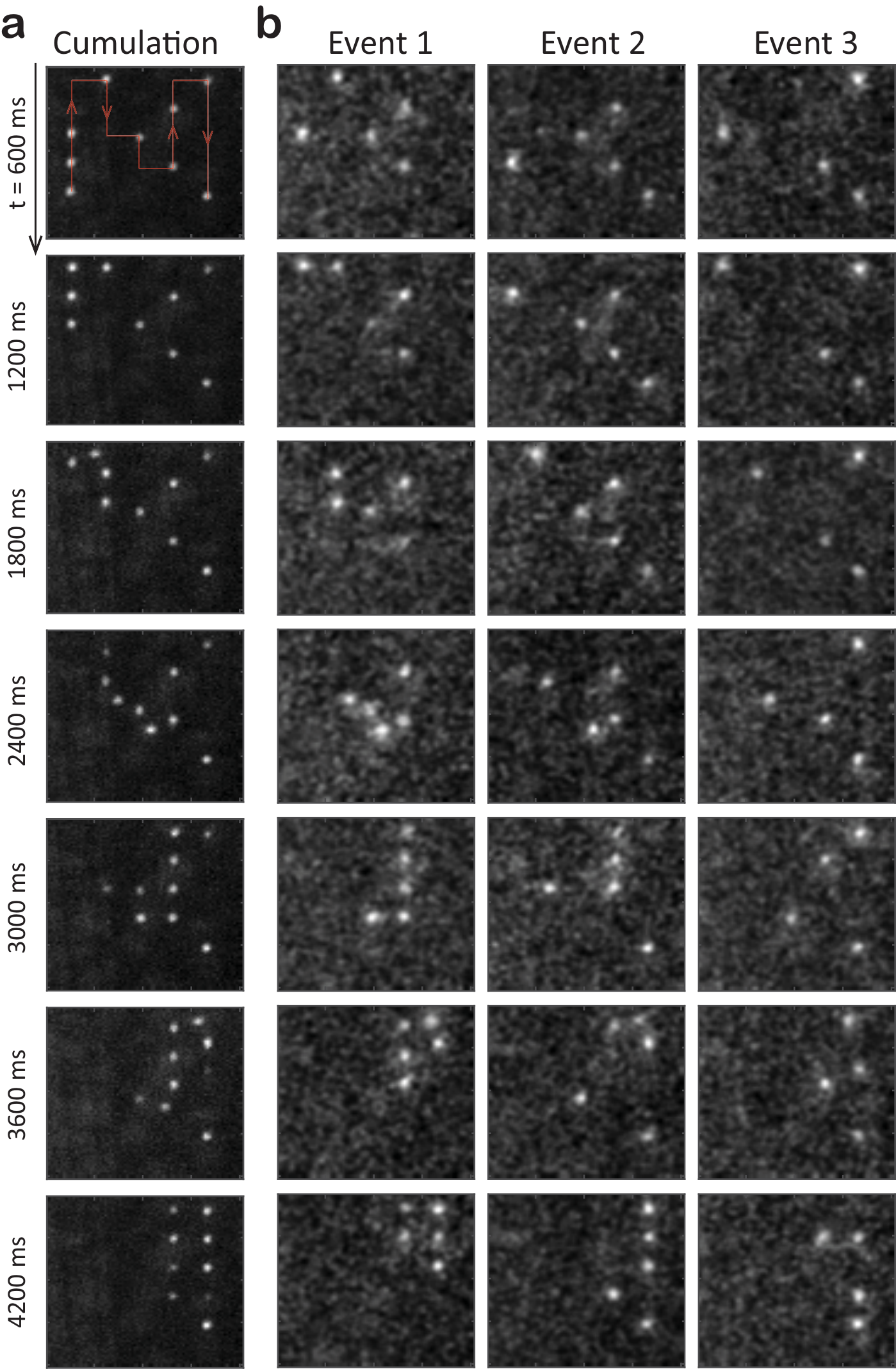}
\caption{ \textbf{Atom movement example.}  \textbf{a}, Transient cumulative images of 500 snapshots showing that nine atoms move the predefined path (red arrows) sequentially. \textbf{b}, Three selected single events ($N \geq 4$). Every snapshot represents the same $26 \times 26~\mu$m$^2$ area. The images are Gaussian filtered for clarity. }
\label{figs2}
\end{figure}

\noindent \textbf{Simultaneous movement of single atoms} Single atoms in an array are moved along the predefined path as shown in Fig.~\ref{figs2}.

\end{document}